# Gaps and Pseudo-gaps at the Mott Quantum Critical Point in the Perovskite Rare Earth Nickelates

S. James Allen[1], Adam J. Hauser[2], Evgeny Mikheev[2], Jack Y. Zhang[2], Nelson E. Moreno[2], Junwoo Son[2], Daniel G. Ouellette[1], James Kally[1], Alex Kozhanov[1] Leon Balents[1,3] and Susanne Stemmer[2]

[1]Department of Physics, University of California, Santa Barbara, California
[2]Materials Department, University of California, Santa Barbara, California
[3]Kavli Institute for Theoretical Physics, University of California, Santa Barbara, California

**Abstract**

We report on tunneling measurements that reveal for the first time the evolution of the quasi-particle state density across the bandwidth controlled Mott metal to insulator transition in the rare earth perovskite nickelates.   In this, a canonical class of transition metal oxides, we study in particular two materials close to the T=0 metal-insulator transition: $NdNiO_3$, an antiferromagnetic insulator, and $LaNiO_3$, a correlated metal. We measure a sharp gap in $NdNiO_3$, which has an insulating ground state, of  $2\Delta \approx 30$ meV. Remarkably, metallic $LaNiO_3$ exhibits a *pseudogap* of the same order that presages the metal insulator transition.  The smallness of both the gap and pseudogap suggests they arise from a common origin: proximity to a quantum critical point at or near the T=0 metal-insulator transition.   It also supports theoretical models of the quantum phase transition in terms of spin and charge instabilities of an itinerant Fermi surface.

The pioneering work of Mott[1] showed that when the strength of Coulomb repulsion between electrons exceeds the electronic bandwidth, strong correlations drive an otherwise metallic state into an insulating one.[2] Despite the significance and age of the topic, the *order* of this "bandwidth tuned" quantum phase transition between distinct ground states remains a subject of active debate.[3-6] Theoretically, the possibility of a second order or continuous T = 0 Mott transition is tied to many interesting phenomena such as marginal Fermi liquid behavior and quantum spin liquid states, so it is highly desirable to address whether the bandwidth tuned Mott transition can be continuous or nearly so in experiment. A continuous quantum phase transition is *quantum critical*, which implies the presence of characteristic energies that vanish on approaching the transition point from either direction (see Fig. 1a.[7] ). On the side for which the ground state is insulating, several obvious energy scales exist, such as the single-particle gap, or the critical temperature for the thermally driven metal-insulator and/or magnetic ordering transition. On the metallic side of the T = 0 Mott transition, however, one has a correlated metallic state without a gap or T>0 transition; finding evidence for a low energy scale associated with Mott physics is more challenging.

Here we present new experimental evidence for a low energy scale on both sides of the T=0 Mott transition in the rare earth perovskite nickelates[8,9] with chemical formula $RNiO_3$, a canonical class of transition metal oxides in which a bandwidth controlled Mott transition[2] is controlled by the size of the rare earth ion R[7]. The rare earth ion effectively tunes the Ni $e_g$ bandwidth and the two materials studied here, $LaNiO_3$ and $NdNiO_3$, sit near the T=0 Mott transition point on the metallic and insulating side, respectively. Our study is based on electron tunneling spectroscopy of epitaxial thin films, which uniquely probes the single particle density of states in correlated electron systems[10] with fine energy resolution over a wide temperature range. We show that in NdNiO$_3$, a sharp gap of $2\Delta \approx 30$ meV develops at low temperature, and moreover broadens but persists as a depression of the tunneling conductance to temperatures above the transition temperature: this is a *pseudogap*, reminiscent of the high temperature cuprate superconductors. Remarkably, we show that LaNiO$_3$ shows a similar pseudogap feature, *despite* its good metallic behavior. To our knowledge this is the first clear observation of a pseudogap feature in the single particle spectrum of a system on the metallic side of a bandwidth controlled Mott transition. We draw attention to the fact that the characteristic temperatures at which these features appear, and the energy (voltage) scale of the features themselves are comparable, and both are much smaller than the e$_g$ bandwidth and Coulomb energy U (~4 eV [11]), intimating a continuous or nearly continuous nature of the T=0 Mott transition in the nickelates.

We begin by characterizing the in plane transport, shown in Figure 1b, in order to correlate our tunneling observations with the onset of the thermal metal-insulator transition. The epitaxial NdNiO$_3$ film is metallic (a positive temperature coefficient of resistance) until it undergoes a sharp, hysteretic transition to an insulating state[12]. The resistance ratio of ~ 10$^6$ measured between the transition temperature and ~ 7.5 K is comparable or larger than the best values reported in the literature, testifying to the high film quality. The transition temperature (T$_{M-I}$ ~ 100 K on cooling) is roughly a factor of 2 smaller than is observed in bulk ceramic samples (T$_{M-I}$ ~ 200K)[13,14]. It has been well documented that the transition temperature for NdNiO$_3$ is strongly affected by substrate induced coherency strains[9,15-22]. It is also established[9,18,23] that the hysteretic transition region in NdNiO$_3$ corresponds to two phase coexistence of insulating and metallic domains. Slow, time dependent transport (time constants of 1000's of seconds) appears below the phase transition.[12] The sharp rise in resistance at

~ 100 K probably marks the temperature at which the insulating region is greater than ~½ the area, and the conducting domains no longer percolate. At ~ 70 K and below, while metallic regions exist, the film is predominantly insulating.[9,18]

The epitaxial LaNiO$_3$ film used for the tunneling spectroscopy exhibited a room temperature resistivity of ~0.15 mΩ-cm and a resistance ratio [R(300 K)/R(7.5K)] of approximately 5 [24]. This is within a factor of 2 of the best ceramic samples, which can exhibit room temperature resistivity as low as 0.1 mΩ-cm with resistance ratios, R(300K)/R(<10K) as large as 10 [13], and is similar to the best published films. Thus, comparison to published resistivity and resistance ratio measurements of bulk ceramics[8,13,14,25] and thin films[26-28], shows that that the films for the tunnel studies are representative of stoichiometric LaNiO$_3$. Broadband infrared conductivity on similar LaNiO$_3$ films, coherently strained to different substrates, have successfully measured correlation driven mass enhancement.[29] The temperature dependent resistivity data displayed in Fig. 1b are obtained on the same films as used for the tunneling junctions.

We now turn to the tunneling measurements on NdNiO$_3$. The tunneling conductance versus bias of a 16.5 nm NdNiO$_3$ film is shown in Fig. 2. The data were taken upon cooling. The tunneling conductance is dramatically altered as the film transitions to the insulating state. However, even before entering the insulating state, the zero bias tunneling conductance is suppressed, an observation to which we will return to later. Immediately below the transition temperature, the tunneling conductance at the Fermi energy, at zero bias, drops by several orders of magnitude while it is strongly enhanced on either side of the gap. The energy resolution of the tunneling probe is ~ $3k_BT$, which corresponds to 10 meV at 50 K. The features in the tunnel conductance between 50-100 K are robust and are consistent with a depletion of the density of states below a many-body correlation gap, which tends to conserve state density by transferring states above the gap. These experimental features are analogous to normal-superconductor tunnel junctions.[30] As the temperature is lowered, the zero-bias conductance continues to fall and the tunneling conductance over the range of voltages that could be measured is strongly suppressed. At the lowest temperatures used in this experiment the thermal smearing is negligible. There are no metallic $NdNiO_3$ inclusions at these temperature, and the tunneling conductance exhibits a sharp turn on at ~ + and - 15 meV. This is a clear and unequivocal indication of a true spectral gap in the insulating state – to our knowledge this is the first such observation in *any* $RNiO_3$ nickelate. The 15 meV magnitude of the gap is comparable in energy to the scale of the critical temperature of the metal-insulator transition. Taking $T_c$ = 100 K (on cooling), and the gap $\Delta$ = 15 meV, one calculates $\Delta/k_BT_c$ = 1.7, which is strikingly close to the BCS value[30] This suggests the metal-insulator transition is driven by excitations of electrons on the scale of the gap, and not much farther from the Fermi energy, and more speculatively that a mean-field description might apply.

The observation of a well-defined gap in the state density in $NdNiO_3$ at low temperatures reported here differs qualitatively from results obtained by infrared conductivity[31,42,43] and photoemission[32,33] measurements. Unlike recent results on $SmNiO_3$[34], a well-defined optical gap has not been reported in $NdNiO_3$. The lack of observation of a hard gap in optics and photoemission in the latter may be attributed to the lower energy resolution in these experiments, or to sub-gap contributions to the optical conductivity from neutral excitations.

It appears that the tunnel conductance at the lowest temperatures has collapsed outside the gap region as well. It is important to determine if the apparent transfer of state density outside the gap persists at the lowest temperatures, and if so what is the energy scale.

Unfortunately at the lowest temperatures, despite the 4-terminal tunnel device used here, the sheet resistance of the NdNiO$_3$ film limits the current and therefore the accessible range of voltage drops across the junction.

The presence of a sharp spectral gap is certainly expected in the clean limit for a Mott insulator like NdNiO$_3$, though the observation is striking and unprecedented in the nickelates. More unexpected is the fact that the depression in the tunneling spectrum is visible even at temperatures *above* the metal-insulator transition. This is shown in Figure 2a and b, and especially in Figure 2b, in which the tunneling conductance is plotted normalized to the curve at T=140 K. One observes a small but clearly visible depression of the tunneling conductance which is progressively more suppressed with reducing temperature between 140 K and 100 K (data is taken on cooling). This is a *pseudogap*, similar to that observed in some other correlated materials, in particular, the under-doped cuprates. In the cuprates, the origin of the pseudogap is hotly debated, with proposed origins including preformed superconductivity, competing charge or spin order, and more exotic scenarios. Here we can at least rule out superconductivity.

A possible prosaic explanation is that the pseudogap might arise from a small concentration of insulating domains in this temperature region. To address this, Fig. 3 displays the zero bias tunneling conductance for falling and rising temperature. It indeed displays hysteresis that follows the hysteresis in the film resistance. The fractional difference in zero bias tunneling conductance for rising and falling temperatures at a given temperature is, however, less than the fractional difference in sheet resistance on cooling and heating. Since there is no evidence for insulating NdNiO$_3$ regions in the sheet resistance on cooling in the same temperature interval that the pseudogap is observed, the coexistence explanation seems unlikely.

The intrinsic nature of the pseudogap in NdNiO$_3$ is further supported by the observation of a very similar effect in LaNiO$_3$, to which we now turn. The tunneling conductance versus bias voltage of the 30 nm LaNiO$_3$ film is shown in Fig. 4. At elevated temperatures the tunneling conductance (d$I$/d$V$) exhibits moderate voltage dependence. Below 100 K, the tunneling conductance is depressed over a relatively narrow range of bias voltage (~ 30 mV). It is displayed most prominently by normalizing the low temperature conductance by the relatively featureless conductance versus voltage at 200K, and then scaling. The depression of the zero bias conductance with decreasing temperature appears to saturate in the temperature range of the experiments. This is of course expected, as LaNiO$_3$ remains metallic at all temperatures, with a non-zero density of states at the Fermi level.

The observation of a pseudogap at all in LaNiO$_3$ is striking. It cannot be attributed to inclusions of a lower temperature phase, as there is no phase transition in LaNiO$_3$. Rather, it should be associated to the correlated nature of the metallic state. Prior evidence for correlations in LaNiO$_3$ comes from angle resolved photoemission (ARPES)[35] and broad band infrared conductivity[29], which have observed substantial mass enhancement, by approximately ~ 3 times the band structure mass. From the heavy effective mass $m^*$, one might naively expect an enhanced rather than suppressed tunneling conductance, since the thermodynamic density of states is inversely proportional to $m^*$, but this may be compensated in the single particle density of states, probed by tunneling, by the quasi-particle renormalization factor $Z$. In simple theories such as the Gutzwiller approximation or Dynamical Mean Field Theory[36], which neglect momentum dependence of the self-energy, these effects cancel, and the tunneling conductance is neither suppressed nor enhanced. The suppression observed here therefore indicates non-trivial momentum dependent self-energy effects.

We comment on prior experiments probing quasi-particle density of states on LaNiO$_3$. Tunneling measurements on LaNiO$_3$ have been published[37-39], but differ both quantitatively and qualitatively from the results reported here. The pseudo-gap feature in the earlier reports is projected on theoretical models of electron-electron interactions acting in concert with disorder. This model produces a pseudo gap that has a density of states at the Fermi energy that behaves as $N(E) = N(0)\left(1 + \left[\left(|E - E_F|\right)/\Delta\right]^{1/2}\right)$ [40]. The correlation gap, $\Delta$, deduced from the published work is as large as ~ 0.58 eV, [38] which is more than an order of magnitude larger than the energy scale of the pseudo-gap shown in Fig. 4. The depression in tunneling conductance cannot be fit to the aforementioned voltage dependence based on disorder scattering in the presence of electron-electron correlations. Thus we exclude the possibility of a disorder-driven gap in our experiments.

In principle photoemission spectroscopy (PES) originating near the Fermi energy can reveal pseudo gap related features in LaNiO$_3$. Recent PES measurements, with a resolution of ~ 120 meV, show suppression of the state density for the thinnest films, less than 10 monolayers, which correlates well with localized electrical transport.[41-43] However, improved resolution would be required for a comparison with the tunneling data.

We conclude with a discussion of the implications of our results. For NdNiO$_3$, we observed for a first time a sharp gap in an insulating nickelate, and determined this gap to be only $2\Delta \sim 30$ meV, an order of magnitude smaller than the prior estimate of 200 meV based on optics.[44] The smallness of the gap presents a challenge to a local picture of the gap formation, in which the natural energy scales are the $e_g$ band width, of order 1-2 eV,[45] the hybridization of the nickel $d$ and oxygen $p$ states, and the on-site Coulomb repulsion $U$ of order ~4 eV .[11,46-48] Instead, the small gap supports theoretical models that are based on an itinerant electron liquid for which the instability arises from a relatively narrow range of states around the Fermi energy.[49,50] The consistency of the $\Delta/k_BT_c$ ratio with the mean-field BCS value adds further value to this proposition and suggests a mean field model based on such a narrow set of states.

The most important aspect of the results is the observation of a pseudogap in the tunneling density of states for $NdNiO_3$ *and* $LaNiO_3$, on *both* sides of the bandwidth-tuned Mott transition in the nickelates. While the pseudogap in $NdNiO_3$ might, on its own, be deemed a precursor to the lower temperature ordering, taken together with the pseudogap in $LaNiO_3$, it seems most natural to associate both with proximity to the T = 0 metal-insulator transition. The energy scales of these pseudogaps, determined both from the voltage scale of the conductance suppression and the onset temperature, are comparable, and are small compared to the microscopic energy scales of the problem. These observations point naturally to the proposition that the T = 0 metal-insulator transition is nearly continuous or quantum critical, and that the pseudogap energy scales probe the (nearly) quantum critical fan emerging from this point.

Given these exciting observations, we envision a wide scope for future studies. Tunneling spectroscopy above the gap may, in a fashion similar to superconductor tunnel junctions, reveal spectroscopic features of excited states and modes at low temperatures that can test microscopic models of the interactions. Criticality of the quantum phase transition can be further addressed by quasi-particle tunneling spectroscopy of the pseudo-gap in strained $LaNiO_3$ films with reduced bandwidth. Transport and infrared conductivity has documented evidence of mass enhancement under tensile strain[29]. On the insulator side of the quantum phase transition, tunneling spectroscopy in $PrNiO_3$ would provide a measure of the insulating gap closer to the quantum phase transition.

**Methods**

16.5-nm-thick epitaxial NdNiO$_3$[12] films and 30-nm-thick epitaxial LaNiO$_3$[24] films were grown on (001) LaAlO$_3$ by r.f. magnetron sputtering. Growth conditions and structural characterization of both are described elsewhere[12,24] and confirm that the films are coherently strained to the substrate. Transmission electron microscopy of a NdNiO$_3$ film grown at the same time as the film in the tunnel device reveal small inclusions of a second phase that may be NiO.

A schematic top view of the 4-terminal tunnel devices is shown in Fig. 5. They were fabricated in the following manner. The 5 mm × 5 mm RNiO$_3$ films were heated in pure oxygen at 600 °C for 30 minutes to ensure oxygen stoichiometry. The tunnel barrier was produced by evaporating ~ 1 nm of Al over the entire film. The Al covered RNiO$_3$ films were heated again for 30 minutes in pure oxygen at 600 °C to form the Al$_2$O$_3$ tunnel barrier and to return the RNiO$_3$ film to a fully oxidized state. .

The RNiO$_3$/Al$_2$O$_3$ film was patterned by scribing through openings in an appropriate mask. Following scribing through mask openings, an insulator was deposited through the same areas to mitigate potential shorts to the Al counter electrodes in the areas physically abraded by the mini-scribe . Al counter electrodes were deposited by thermal evaporation through another shadow mask, resulting in 4 tunnel junctions on each 5 mm square chip. Electrical connection was made by In soldering to the ends of the inscribed strip of the RNiO$_3$ and the ends of an Al counter electrode.

The 4-terminal device is particularly important for the following reasons. Electrical resistivity measurements are made on a processed film to ensure that the characteristics of the film under test in the tunnel junction can be identified with the well characterized resistive transition. Further as the NdNiO$_3$ undergoes the metal insulator transition, the resistance rises dramatically and the 4-terminal measurements mitigate complications that would mask the tunnel conductance by the series resistance of the film. Electrical measurements were made with a Keithley source-meter while the test device was mounted on the cold finger of a closed cycle refrigerator that could reach 7.5 K.

**Acknowledgements**

Work by A.J.H. and E. M. was supported in part by FAME, one of six centers of STARnet, a Semiconductor Research Corporation program sponsored by MARCO and DARPA. A.J.H. also acknowledges support through an Elings Prize Fellowship of the California Nanosystems Institute at University of California, Santa Barbara. S.S., S.J.A, D.G.O., L.B., J. S., E.M., J.K. and A.K. acknowledge support by a MURI program of the Army Research Office (Grant # W911-NF-09-1-0398). D.G.O. and N.M. acknowledge support by the UCSB MRL (supported by the MRSEC Program of the NSF under Award No. DMR 1121053). This work also made use of the Central Facilities of the UCSB MRL. N.M. was a CAMP RISE Intern and was supported by NSF via UC Irvine (NSF HRD-1102531-California Louis Stokes Alliance for Minority Participation Phase V). The work also made use of the UCSB Nanofabrication Facility, a part of the NSF-funded NNIN network.

**Authors contributions**

A.J.H., N. M., and E. M. grew the material. S.J.A., D.G.O. J.K. and A.K fabricated test devices. S.J.A. carried out electrical measurements. J.Y.Z. determined the structure of the films. S.J.A., L.B., S.S., and A.J.H. wrote the manuscript. All authors participated in interpreting the experimental results and editing the manuscript.

**Competing financial interest.**

The authors declare no competing financial interest.

Figure 1. a) The phase diagram of the RNiO$_3$'s near the quantum phase transition. The changing rare earth ionic radius is thought to control the band width, W, and its ratio to the on site coulomb repulsion, U, tunes the system through the transition. The line marking the transition to the insulating state is taken from Torrance et al. [7], but with the temperature axis scaled so that the transition temperature coincides with that observed in the films under investigation. b) Sheet resistance of a 16.5 nm thick film of NdNiO$_3$ on LaAlO$_3$ versus temperature and the sheet resistance of a 30. nm thick film of LaNiO$_3$ on LaAlO$_3$ versus temperature. The data in this figure is obtained from the films that are integral to the tunnel junctions.

Figure 2. a) Tunneling conductance (logarithmic scale) for NdNiO$_3$ versus voltage at various temperatures (cooling). b) Tunneling conductance normalized to the conductance at 140 K. The linear scale exposes the increase in tunneling conductance at elevated voltages, outside the evolving energy gap, as the film becomes insulating. The tunneling conductance completely collapses at the lowest temperatures.

Figure 3. Zero bias tunneling conductance versus temperature, cooling and heating, showing hysteresis that mirrors the sheet conductance in Fig. 1b: Filled symbols cooling, open symbols warming.

Figure 4. a) Tunneling conductance for LaNiO$_3$ versus voltage at various temperatures. b) Normalized tunneling conductance versus voltage at various temperatures. The conductance is first normalized to the conductance vs voltage at 200K and then scaled to be 1. at -75 meV.

Figure 5. A schematic of the 5 mm x 5 mm chip that presents tunnel junctions for a 4-terminal measurement of conductance. It also allows a two-terminal measurement of the conductance of the film that is one electrode of the tunnel device.

Figure 1

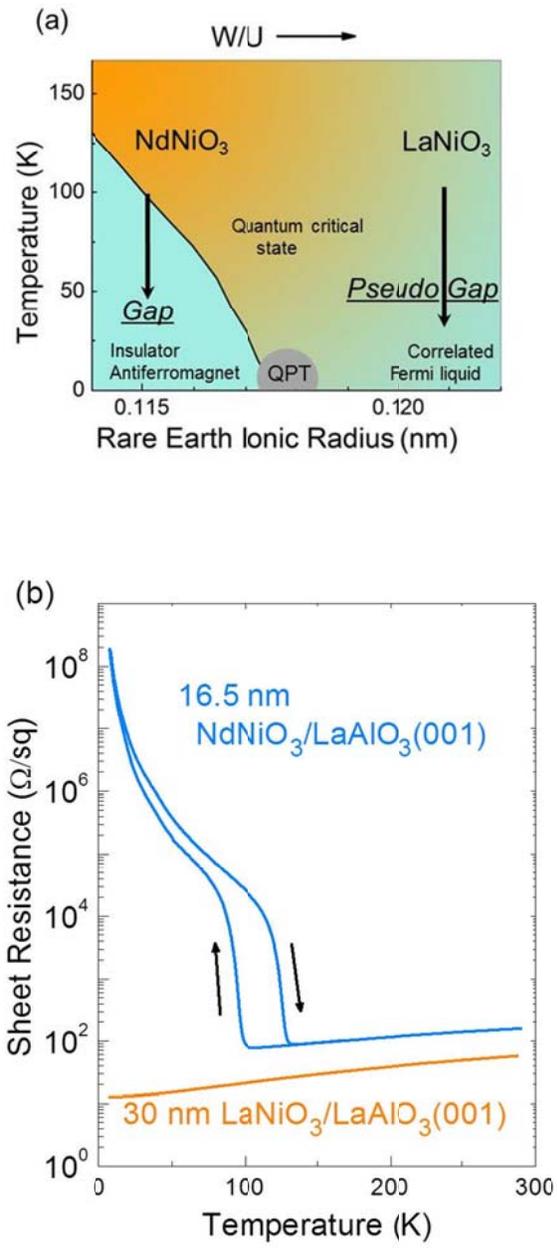

Figure 2

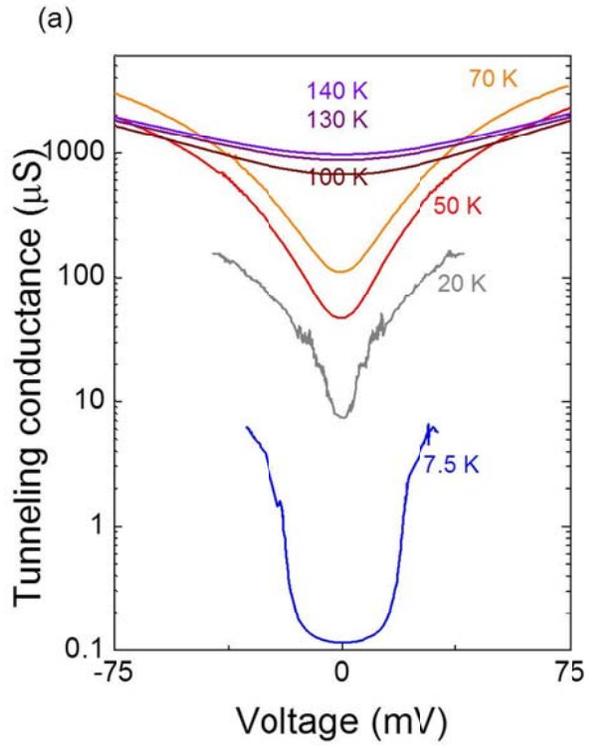

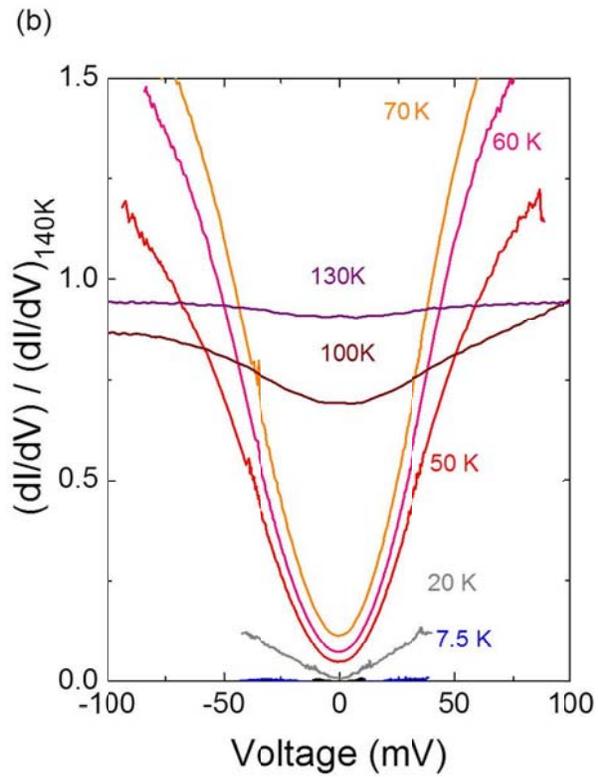

Figure 3

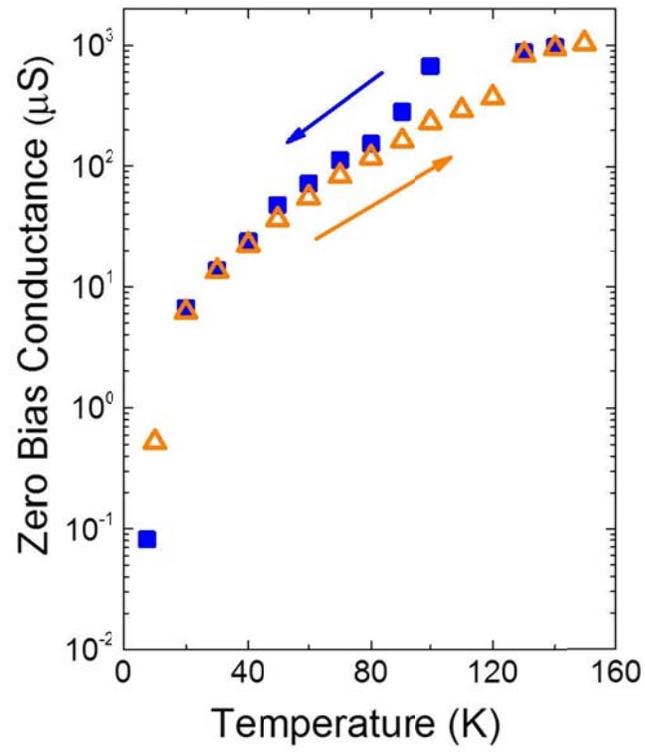

Figure 4

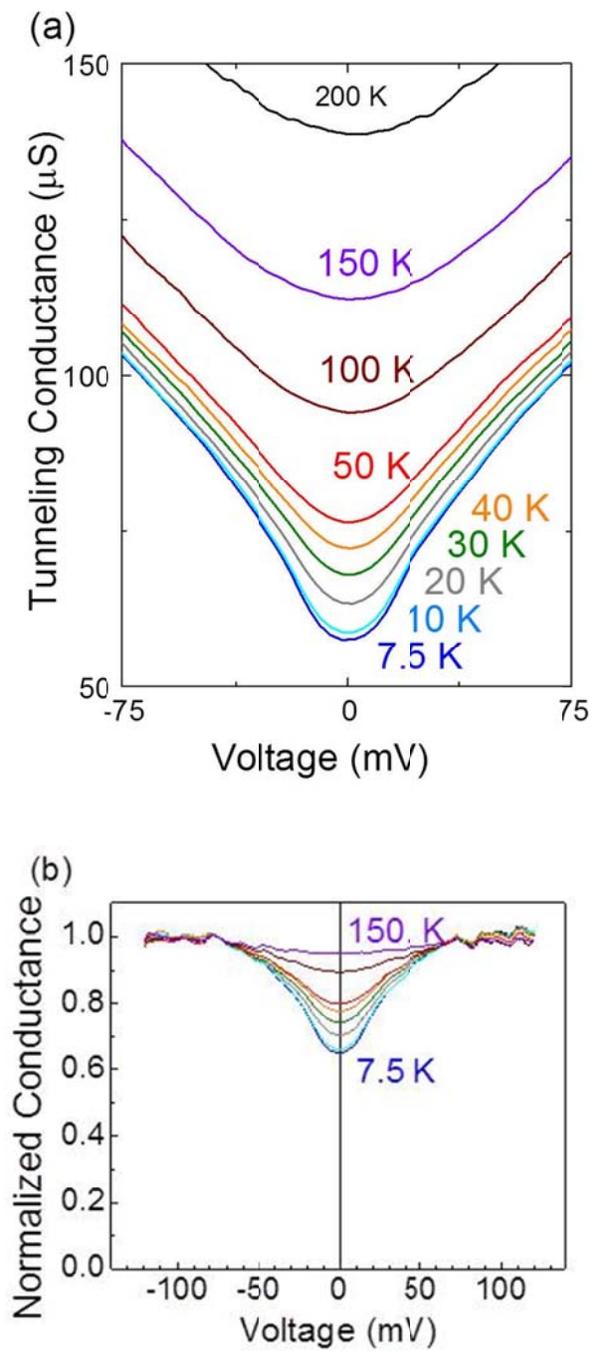

Figure 5

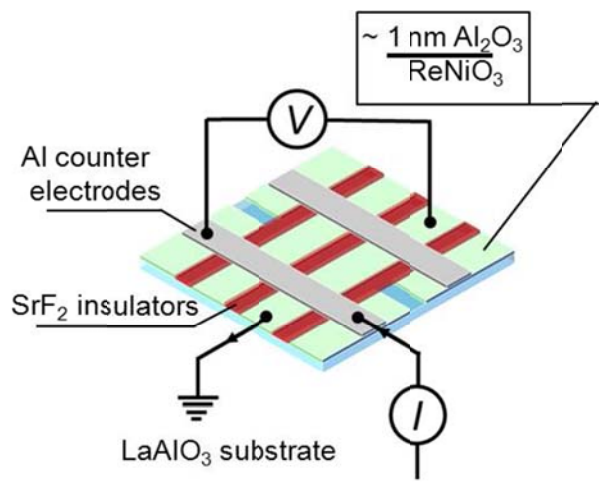